\documentclass[10pt,conference]{IEEEtran}
\IEEEoverridecommandlockouts
\usepackage{cite}
\usepackage{amsmath,amssymb,amsfonts}
\usepackage{algorithmic}
\usepackage{graphicx}
\usepackage{textcomp}
\usepackage{xcolor}
\usepackage{booktabs}
\usepackage{soul}
\usepackage{url}
\usepackage{listings}
\usepackage{tikz}
\usepackage{colortbl}
\usepackage[hidelinks]{hyperref}

\usepackage{listings}
\usepackage{xcolor}
\usepackage{tikz}

\newcommand{\codecallout}[1]{%
    \tikz[baseline=(n.base)]{
        \node[
            circle,
            fill=black,
            text=white,
            inner sep=1.2pt,
            minimum size=4mm,
            font=\scriptsize\bfseries
        ] (n) {#1};
    }%
}

\newcommand{\codecalloutbox}[1]{%
    \makebox[5mm][l]{\codecallout{#1}}%
}

\lstdefinestyle{ragitcode}{
    language=Python,
    basicstyle=\ttfamily\scriptsize,
    frame=single,
    rulecolor=\color{black!40},
    backgroundcolor=\color{black!3},
    columns=fullflexible,
    keepspaces=true,
    showstringspaces=false,
    breaklines=true,
    xleftmargin=3pt,
    xrightmargin=3pt,
    framexleftmargin=2pt,
    framexrightmargin=2pt,
    aboveskip=2pt,
    belowskip=2pt,
    escapeinside={(*@}{@*)},
    emph={
        initialize_collection,
        ParsingConfig,
        ChunkingConfig,
        IndexingConfig,
        RetrievalConfig,
        EvaluationConfig,
        ExperimentConfig,
        run_benchmark
    },
    emphstyle=\bfseries
}

\def\BibTeX{{\rm B\kern-.05em{\sc i\kern-.025em b}\kern-.08em
    T\kern-.1667em\lower.7ex\hbox{E}\kern-.125emX}}
\begin{document}

\title{RAGit: Turning Document Preprocessing into an Experimental Variable for RAG{}
}



\author{
\IEEEauthorblockN{
Omar El Bachyr\textsuperscript{*},
Saad Ezzini\textsuperscript{\ddag},
Tegawendé F. Bissyandé\textsuperscript{*},
Jacques Klein\textsuperscript{*},
}

\IEEEauthorblockA{
\textsuperscript{*}
University of Luxembourg, Luxembourg\\
\{omar.elbachyr, tegawende.bissyande, jacques.klein\}@uni.lu
}

\IEEEauthorblockA{
\textsuperscript{\ddag}
King Fahd University of Petroleum and Minerals, Dhahran, Saudi Arabia\\
saad.ezzini@kfupm.edu.sa
}
}

\maketitle

\begin{abstract}
Document preprocessing choices can substantially affect retrieval in Retrieval-Augmented Generation (RAG) pipelines, yet they are often selected heuristically and are difficult to compare under consistent conditions. We present \texttt{RAGit}, an open-source experimentation workbench for the systematic and reproducible evaluation of document preprocessing choices through downstream retrieval. \texttt{RAGit} represents parsing, chunking, indexing, retrieval, and evaluation as configurable experimental stages, enabling controlled comparisons while the remaining downstream settings are kept fixed. It preserves provenance from retrieved chunks to source pages and maps heterogeneous retrieval units to a common page-level ranking for evaluation. The current implementation supports multiple parsers and chunking strategies, dense and late-interaction retrieval, reusable intermediate artifacts, and multi-configuration benchmarking. We demonstrate \texttt{RAGit} through controlled experiments comparing alternative preprocessing configurations under a fixed retrieval setup, illustrating how preprocessing can be treated as an explicit experimental variable. \texttt{RAGit} and its documentation are \href{https://github.com/OmarElbachyr/ragit}{publicly available on GitHub}. A demonstration video is available at \url{https://youtu.be/tsst-Bhfz4I}.
\end{abstract}


\section{Introduction}
\begin{figure*}[!t]
    \centering
    \scalebox{0.78}{%
        \includegraphics[width=\textwidth]{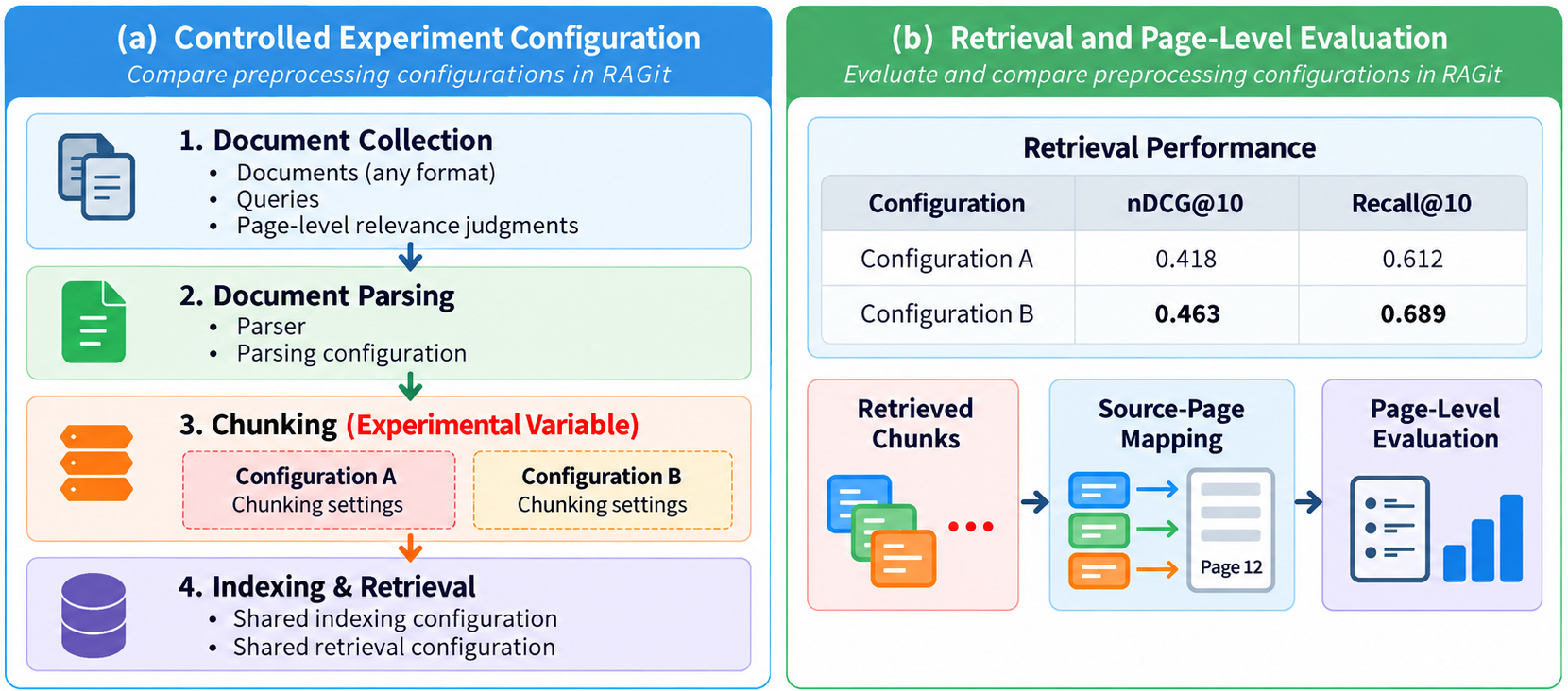}
    }
    \caption{Overview of a controlled \texttt{RAGit} experiment. (a) Alternative preprocessing configurations are compared while the downstream indexing, retrieval, and evaluation settings remain fixed. (b) Retrieved chunks are mapped to their source pages for page-level evaluation, and the resulting metrics are summarized for comparison.}
    \label{fig:ragit-usage}
\end{figure*}
Retrieval-Augmented Generation (RAG) increasingly operates over complex
documents such as technical reports, scientific articles, manuals, scanned
documents, and slide decks. Before retrieval, these documents must be parsed and split into suitable retrieval units. This introduces several consequential choices, including the parser and its configuration, the chunking strategy, chunk size, and overlap. For visually rich documents, where text, tables, figures, and layout interact, these choices can substantially change the content exposed to the retriever. Nevertheless, preprocessing is often treated as an implementation detail rather than as an experimental variable.

Recent studies have started to quantify the impact of preprocessing choices on retrieval and downstream RAG performance. One study systematically evaluates PDF parsing and chunking choices in financial question answering~\cite{el2026empirical}, while another compares PDF conversion and chunking configurations through downstream RAG evaluation~\cite{marques2026pdf}.. These studies show that preprocessing can materially affect downstream performance, but their evaluation pipelines are tailored to individual studies rather than designed as reusable tools for systematically comparing and extending preprocessing configurations.

General RAG experimentation frameworks address adjacent problems. AutoRAG~\cite{kim2024autorag} explores combinations of RAG components, Rankify~\cite{abdallah2026rankify} supports unified retrieval and reranking experimentation, and RAGChecker~\cite{ru2024ragchecker} provides fine-grained RAG evaluation. Their focus, however, is broader pipeline optimization or diagnosis rather than controlled evaluation of document preprocessing. A particular challenge is comparing preprocessing configurations that produce different retrieval units, such as complete pages, smaller chunks, or cross-page chunks, while benchmark relevance judgments remain defined at the page level.

We present \texttt{RAGit}, an open-source experimentation workbench for the systematic and reproducible evaluation of document preprocessing choices for retrieval in RAG pipelines. \texttt{RAGit} represents parsing, chunking, indexing, retrieval, and evaluation as explicit experimental configurations, allowing preprocessing choices to be varied while downstream conditions remain fixed. It preserves provenance from retrieved chunks to their originating source pages, allowing heterogeneous retrieval units to be mapped and aggregated into a common page-level ranking for evaluation. The resulting workflow is illustrated
in Figure~\ref{fig:ragit-usage}.

\texttt{RAGit} is designed for both researchers studying the effect of preprocessing on retrieval and practitioners comparing alternative document-processing configurations before integrating them into RAG systems. The current implementation supports multiple parsers and chunking strategies, dense and late-interaction retrieval, reusable intermediate artifacts, and controlled comparison of multiple experimental configurations.

The contributions of this tool demonstration are:

\begin{itemize}
    \item We introduce \texttt{RAGit}, an open-source workbench for configurable
    document preprocessing and retrieval experiments.

    \item We provide provenance-preserving mapping from retrieved chunks to
    source pages, separating retrieval and evaluation units.

    \item We support reproducible comparison of multiple configurations under
    shared downstream settings.
\end{itemize}

The remainder of the paper presents the \texttt{RAGit} architecture, an
illustrative usage scenario, and its empirical validation.

\section{RAGit Tool Overview and Architecture}
\label{sec:ragit}

\texttt{RAGit} is an experimentation workbench for systematically comparing document preprocessing and retrieval choices in RAG pipelines. It represents the parser and its configuration, chunking strategy and parameters, and retrieval setup as explicit experimental variables, enabling controlled comparisons while keeping the remaining conditions fixed.

A \texttt{RAGit} experiment follows five stages: \emph{parsing}, \emph{chunking}, \emph{indexing}, \emph{retrieval}, and \emph{evaluation}. Source pages are first converted into a normalized representation, then split into retrieval chunks. The chunks are indexed and retrieved for each query, after which their scores are mapped back to source pages and aggregated into a page-level ranking for retrieval evaluation. The overall workflow is illustrated in Figure~\ref{fig:ragit-usage}.

\subsection{Configurable Document Preprocessing}

\textbf{Parsing.} \texttt{RAGit} decouples parser-specific document processing from the rest of the pipeline through a normalized page representation. Each source page is represented internally as a \texttt{ParsedPage}, with its content exposed as plain text or Markdown depending on the selected parser and configuration. This allows different parsing backends to feed the same chunking, indexing, retrieval, and evaluation pipeline.
The supported parsers and their main capabilities are summarized in Table~\ref{tab:ragit-parsers}. In addition to conventional text extraction, some parsers support OCR, parsing complete documents with a VLM, or generating semantic descriptions of detected pictures.

\begin{table}[!t]
    \caption{Built-in parsers in \texttt{RAGit}.}
    \label{tab:ragit-parsers}
    \centering
    \footnotesize
    \setlength{\tabcolsep}{3pt}
    \begin{tabular}{lll}
        \hline
        \textbf{Parser} & \textbf{Output} & \textbf{Optional capabilities} \\
        \hline
        PyMuPDF4LLM~\cite{pymupdf4llm} & Text, Markdown & OCR \\
        Docling~\cite{auer2024docling} & Text, Markdown & OCR, VLM, picture descriptions \\
        Marker~\cite{marker} & Markdown & OCR, picture descriptions \\
        \hline
    \end{tabular}
\end{table}

Once a source document has been parsed, the resulting page-level content is persisted and reused for the same parser configuration. Missing pages are processed as needed, while a different configuration produces a separate parsing artifact. This avoids repeated processing in computationally intensive pipelines involving layout processing, OCR, VLM-based document parsing, or picture description.

\textbf{Chunking.} Chunking is applied after parsing and is kept independent from the parser. \texttt{RAGit} reconstructs each document by concatenating the parsed content of its pages into a continuous text stream while preserving page boundaries and provenance. This allows chunks to cross page boundaries and capture context that would otherwise be cut off at page transitions.

\texttt{RAGit} supports several chunking strategies, summarized in Table~\ref{tab:ragit-chunkers}. It provides a native strategy that treats each source page as a single chunk, while the remaining built-in strategies use implementations from the Chonkie library~\cite{chonkie2025}. Configuration is strategy-specific; for example, fixed-size strategies control chunk size and overlap between consecutive chunks, while semantic strategies can use an embedding model and similarity threshold. \texttt{RAGit} preserves the links between resulting chunks and their source pages for later aggregation and evaluation. The complete set of options is provided in the \texttt{RAGit} documentation.

\begin{table}[!t]
    \caption{Chunking strategies supported by \texttt{RAGit}, with representative options.}
    \label{tab:ragit-chunkers}
    \centering
    \footnotesize
    \setlength{\tabcolsep}{3pt}
    \begin{tabular}{lll}
        \hline
        \textbf{Strategy} & \textbf{Principle} & \textbf{Example options} \\
        \hline
        Page      & Source page              & None \\
        Token     & Token-based              & Size, overlap, tokenizer \\
        Sentence  & Sentence-preserving      & Size, overlap, tokenizer \\
        Recursive & Hierarchical             & Size, rules \\
        Fast      & Byte-based               & Size, delimiters \\
        Semantic  & Similarity-based         & Size, threshold, embedding model \\
        Late      & Context-aware            & Size, embedding model \\
        Neural    & Semantic-shift detection & Neural model \\
        \hline
    \end{tabular}
\end{table}

\subsection{Retrieval and Evaluation}

\textbf{Retrieval.} The chunks produced in the previous stage are indexed and searched using a configurable retrieval setup. \texttt{RAGit} currently supports two neural retrieval paradigms: dense retrieval and late interaction. Dense retrieval encodes queries and chunks as single vectors and indexes them with FAISS~\cite{karpukhin-etal-2020-dense,douze2024faiss}, while late interaction preserves token-level representations and is supported through FastPlaid, following the ColBERT paradigm~\cite{khattab2020colbert,fastplaid2025}.

Because retrieval is configured independently, the same parsed and chunked content can be compared across retrieval models, or preprocessing variants can be evaluated under a fixed retriever.

\textbf{Evaluation.} To evaluate retrieval, \texttt{RAGit} uses a set of queries $\mathcal{Q}$ with page-level relevance judgments over a document collection $\mathcal{D}$. Each retrieved chunk $c$ retains its provenance $\pi(c)$ to the source page or pages that contributed to its content. Given a chunk score $s(q,c)$, \texttt{RAGit} aggregates the scores of all chunks associated with a page $p$ into a page-level score:
\begin{equation}
S(q,p) = A\left(\left\{s(q,c)\mid p \in \pi(c)\right\}\right).
\end{equation}

Maximum aggregation is used by default, while mean, sum, and user-defined aggregation functions are also supported. This produces a common page-level ranking independently of the retrieval unit or chunking strategy.

We retrieve the top 100 chunks and evaluate the resulting page ranking at a fixed cutoff of 10. We report Recall and nDCG as complementary retrieval metrics: Recall measures how much relevant evidence is retrieved, while nDCG also accounts for ranking position.

\subsection{Experiment Management and Extensibility}

\textbf{Experiment Management.} \texttt{RAGit} stores the configuration and outputs of each pipeline stage alongside the corresponding document collection. Intermediate artifacts are identified by their configuration and reused when possible, avoiding recomputation of unchanged stages and supporting reproducibility.

For systematic comparison, \texttt{RAGit} provides a benchmarking interface that groups several labeled experiment configurations under the same evaluation setup. Each experiment keeps its own stage artifacts and results, while the benchmark produces a common summary table for comparing their retrieval performance.

\textbf{Extensibility.} \texttt{RAGit} allows new preprocessing methods to be integrated without requiring separate indexing, retrieval, or evaluation logic. User-defined chunking strategies can be added as lightweight functions, explicit chunk spans with precise boundaries and metadata, or complete chunker classes. New parsers can also be registered through the parser interface, provided that they produce the normalized \texttt{ParsedPage} output and preserve document and page provenance.

Retrieval models are configured independently and compatible embedding models can be loaded directly from Hugging Face Hub\footnote{\url{https://huggingface.co/models}}, making it straightforward to compare different dense or late-interaction models.

\begin{figure}[!t]
    \centering
    \begin{minipage}{0.98\columnwidth}

\begin{lstlisting}[style=ragitcode]
# Documents, queries, and page-level qrels
collection = initialize_collection(...)

parsing = ParsingConfig(   (*@\codecalloutbox{1}@*)
    parser_name="pymupdf4llm",
    output_format="markdown",
    options={"use_ocr": True})

indexing = IndexingConfig(   (*@\codecalloutbox{2}@*)
    index_type="dense",
    model_name= "Qwen/Qwen3-Embedding-0.6B")

retrieval = RetrievalConfig(
    top_k=100, aggregation="max")

evaluation = EvaluationConfig(
    metrics=["nDCG", "Recall"], k=[10])

chunking_512 = ChunkingConfig(
    chunker_name="token",
    options={"tokenizer": "character",
             "chunk_size": 512})

# Same configuration, with chunk_size=1024
chunking_1024 = ChunkingConfig(...)

# Same pipeline; only chunk size differs.
experiments = {   (*@\codecalloutbox{3}@*)
    "token-512": ExperimentConfig(
        parsing=parsing,
        chunking=chunking_512,
        indexing=indexing,
        retrieval=retrieval,
        evaluation=evaluation),

    "token-1024": ExperimentConfig(
        parsing=parsing,
        chunking=chunking_1024,
        ...)
}

result = run_benchmark(   (*@\codecalloutbox{4}@*)
    collection=collection, experiments=experiments)

print(result.results_table)
\end{lstlisting}

    \end{minipage}

    \caption{Illustrative use of the \texttt{RAGit} benchmarking API:
    \textbf{1} configure document parsing;
    \textbf{2} define the shared indexing, retrieval, and evaluation settings;
    \textbf{3} define two experiments that share the same pipeline while varying the chunk size; and
    \textbf{4} execute the benchmark and display its comparison table.}
    \label{fig:ragit-benchmark-code}
\end{figure}

\section{An Illustrating Usage Scenario}
To illustrate a typical use of \texttt{RAGit}, we compare two configurations over the same document collection using the same token-based chunking strategy but different chunk sizes. Both experiments share the same parser, indexing method, retrieval settings, and evaluation protocol. Figure~\ref{fig:ragit-benchmark-code} shows how this controlled comparison is configured through the \texttt{RAGit} benchmarking API.

\codecallout{1} configures document parsing using PyMuPDF4LLM with Markdown output and OCR enabled. \codecallout{2} defines the shared indexing, retrieval, and evaluation settings, using Qwen3-Embedding 0.6B, top-100 retrieval, maximum page-score aggregation, and evaluation with nDCG@10 and Recall@10. The two configurations use the same token-based chunking strategy with different chunk sizes. \codecallout{3} defines the two \texttt{ExperimentConfig} instances, while \codecallout{4} executes the benchmark and compares them under the same evaluation setup. The resulting comparison table is available through \texttt{result.results\_table}, with an example illustrated in Figure~\ref{fig:ragit-usage}(b).

\section{Validation and Evidence}
\begin{table}[t]
    \caption{Retrieval effectiveness comparison of chunking strategies.}
    \label{tab:chunking-model-scale}
    \centering
    \footnotesize
    \setlength{\tabcolsep}{4pt}
    \begin{tabular}{lllll}
        \toprule
        & \multicolumn{2}{l}{\textbf{0.6B}}
        & \multicolumn{2}{l}{\textbf{4B}} \\
        \cmidrule(lr){2-3}
        \cmidrule(lr){4-5}
        \textbf{Chunking} &
        \textbf{nDCG@10} &
        \textbf{Recall@10} &
        \textbf{nDCG@10} &
        \textbf{Recall@10} \\
        \midrule
        \rowcolor{gray!15}
        Page
        & 64.48
        & 67.63
        & 70.48
        & 72.86 \\

        Token-512
        & 66.78 {\scriptsize$\uparrow$2.30}
        & 71.75 {\scriptsize$\uparrow$4.12}
        & 71.24 {\scriptsize$\uparrow$0.76}
        & 76.64 {\scriptsize$\uparrow$3.78} \\

        Recursive-512
        & \textbf{67.55} {\scriptsize$\uparrow$3.07}
        & \textbf{72.55} {\scriptsize$\uparrow$4.92}
        & \textbf{72.23} {\scriptsize$\uparrow$1.75}
        & \textbf{77.52} {\scriptsize$\uparrow$4.66} \\
        \bottomrule
    \end{tabular}
\end{table}
We validate \texttt{RAGit} on the ViDoRe-v3 Computer Science collection~\cite{loison2026vidore} through a controlled comparison of Page, Token-512, and Recursive-512 chunking using Qwen3-Embedding (0.6B and 4B). Documents are parsed with PyMuPDF4LLM into Markdown, while the remaining retrieval settings are kept fixed. Results are reported as percentages using nDCG@10 and Recall@10, with arrows indicating absolute percentage-point changes relative to the Page baseline.
Table~\ref{tab:chunking-model-scale} shows that both chunk-based
configurations improve over the Page baseline for both model sizes, with
Recursive-512 obtaining the highest scores in each case. Relative to Page,
Recursive-512 improves nDCG@10 and Recall@10 by 3.07 and 4.92 points with
the 0.6B model, and by 1.75 and 4.66 points with the 4B model.

Since Recursive-512 obtains the highest scores in the main comparison, we use the 4B model to examine the sensitivity of recursive chunking to chunk size. Table~\ref{tab:recursive-size-sensitivity} compares sizes 256, 512, and 1024 while keeping all other settings fixed, with arrows indicating absolute percentage-point changes relative to size 512. Sizes 256 and 512 perform similarly, whereas increasing the size to 1024 decreases nDCG@10 by 5.27 points and Recall@10 by 4.11 points. Together, these experiments illustrate how \texttt{RAGit} supports controlled comparisons of preprocessing choices and quantifies their effect under a fixed retrieval setup.

\begin{table}[t]
    \caption{Retrieval effectiveness comparison across recursive chunk sizes with Qwen3-Embedding-4B.}
    \label{tab:recursive-size-sensitivity}
    \centering
    \footnotesize
    \setlength{\tabcolsep}{4pt}
    \begin{tabular}{lll}
        \toprule
        \textbf{Chunk size} &
        \textbf{nDCG@10} &
        \textbf{Recall@10} \\
        \midrule
        \rowcolor{gray!15}
        512
        & 72.23
        & 77.52 \\

        256
        & \textbf{72.27} {\scriptsize$\uparrow$0.04}
        & \textbf{77.95} {\scriptsize$\uparrow$0.43} \\

        1024
        & 66.96 {\scriptsize$\downarrow$5.27}
        & 73.41 {\scriptsize$\downarrow$4.11} \\
        \bottomrule
    \end{tabular}
\end{table}

\section{Conclusion and Future Work}
We presented \texttt{RAGit}, an experimentation workbench for systematic and reproducible evaluation of document preprocessing choices for retrieval in RAG pipelines. \texttt{RAGit} benefits both researchers and practitioners by providing a common environment for configuring and comparing document preprocessing and retrieval configurations. For researchers, it supports controlled experimentation with preprocessing and retrieval choices; for practitioners, it facilitates the comparison of alternative configurations before their integration into RAG systems. Future work will extend \texttt{RAGit} with additional pipeline components, including rerankers and generation, enabling evaluation of both retrieval effectiveness and generated-answer quality. We also plan to broaden retrieval support to sparse embedding and multimodal models, expand the available parsing and chunking strategies, and provide richer experiment analysis and management capabilities.


\bibliographystyle{IEEEtran}
\bibliography{references}

\end{document}